\documentclass[twocolumn,prl,superscriptaddress]{revtex4-1}

\usepackage{graphicx}
\usepackage{amssymb}
\usepackage{amsmath}
\usepackage{color}

\renewcommand{\vec}[1]{\mathbf{#1}}
\renewcommand{\Re}{\mathop{\mathrm{Re}}}
\renewcommand{\Im}{\mathop{\mathrm{Im}}}
\newcommand{\ep}{\epsilon}

\begin{document}

\title{Cyclotron motion in the vicinity of
a Lifshitz transition in graphite}

\author{M. Orlita}
\affiliation{Laboratoire National des Champs Magn\'etiques Intenses, CNRS-UJF-UPS-INSA, Grenoble, France}
\affiliation{Charles University, Faculty of Mathematics and
Physics, Ke Karlovu 5, 121 16 Praha 2, Czech
Republic}

\author{P. Neugebauer}
\affiliation{Laboratoire National des Champs Magn\'etiques Intenses, CNRS-UJF-UPS-INSA, Grenoble, France}
\altaffiliation[Present address:]{Institut f\"{u}r Physikalische und Theoretische Chemie, J.W. Goethe-Universit\"{a}t Frankfurt, Germany}

\author{C. Faugeras}
\affiliation{Laboratoire National des Champs Magn\'etiques Intenses, CNRS-UJF-UPS-INSA, Grenoble, France}

\author{A.-L. Barra}
\affiliation{Laboratoire National des Champs Magn\'etiques Intenses, CNRS-UJF-UPS-INSA, Grenoble, France}

\author{M. Potemski}
\affiliation{Laboratoire National des Champs Magn\'etiques Intenses, CNRS-UJF-UPS-INSA, Grenoble, France}

\author{F. M. D. Pellegrino}
\affiliation{Dipartimento di Fisica e Astronomia, Universit\`a di Catania,
Via S. Sofia 64, I-95123 Catania, Italy}
\affiliation{CNISM, UdR Catania, I-95123 Catania, Italy}

\author{D. M. Basko}
\affiliation{Universit\'{e} Grenoble 1/CNRS, LPMMC UMR 5493,
B.P. 166, 38042 Grenoble, France}

\begin{abstract}
Graphite, a model (semi)metal with trigonally warped bands, is
investigated with magneto-absorption experiment and viewed as an
electronic system in the vicinity of the Lifshitz transition.
A characteristic pattern of up to twenty cyclotron resonance
harmonics has been observed. This large number of resonances, their
relative strengths and characteristic shapes trace the universal properties of the electronic
states near a separatrix in the momentum space.
Quantum-mechanical perturbative methods with respect to the trigonal warping term hardly
describe the data which are, on the other hand, fairly well reproduced within a quasi-classical
approach and conventional band structure model. Trigonal symmetry is preserved in graphite in contrast to a
similar system, bilayer graphene.
\end{abstract}

\maketitle

Lifshitz transition~\cite{Lifshitz1960} (also known as electronic
topological transition) is a change in the Fermi surface topology
occurring upon a continuous change of some external parameter,
such as pressure~\cite{Pressure}, magnetic field~\cite{Field} or,
most naturally, doping~\cite{Lifshitz-doping}. This transition
does not involve a symmetry breaking, alike conventional phase
transitions of the Landau type, but still leads to observable
singularities in thermodynamics, electron transport, sound
propagation, and magnetic response of metals~\cite{Blanter1994}.
Saddle points in electronic dispersion, often apparent in complex
metals, have only recently been visualized with the spectroscopy
method of angle-resolved photoemission~\cite{Liu2010}. In this
Letter, we show how the proximity to a Lifshitz transition
manifests itself in cyclotron resonance (CR) absorption
experiments on graphite, a model system with saddle points due to
the trigonal warping of electronic bands~\cite{Brandt1988}.

Classically, CR can be understood from the equation of motion for
an electron in a
 magnetic field~$\vec{B}$~\cite{Abrikosov}:
\begin{equation}\label{Newton=}
{d\vec{p}}/{dt}=({e}/{c})\left[\vec{v}\times\vec{B}\right],
%+e\vec{E}(t)
\end{equation}
where $\vec{p}=\hbar\vec{k}$ is the electron quasi-momentum,
$e=-|e|$ the electron charge, and
$\vec{v}=\partial\ep(\vec{p})/\partial\vec{p}$ is the electron
velocity, determined by the dispersion $\ep(\vec{p})$. Since both
the energy $\ep$) and the momentum component~$p_z$ along~$\vec{B}$
are conserved, the motion occurs along cyclotron orbits in the
$(p_x,p_y)$ plane, determined by the condition
$\ep(p_x,p_y,p_z)=\mathrm{const}$. This motion is periodic, and
its period, $2\pi/\omega_c$,being proportional to the cyclotron mass, %$m_c=|eB/\omega_c|$,
defines the cyclotron frequency $\omega_c=\omega_c(\ep,p_z)$.
When an electric field, oscillating at frequency~$\omega$, is
applied, the electron can absorb energy. Absorption becomes
resonant when the perturbation frequency $\omega$ matches the
cyclotron frequency~$\omega_c$ or its integer multiple.

In good metals, the incoming radiation is efficiently screened and
penetrates the sample only within a thin skin layer. CR absorption
is then a surface effect, observed mainly when the magnetic field
is parallel to the surface~\cite{AzbelKaner1956,Khaikin1961}. This makes CR
for good metals a less efficient tool for probing the Fermi
surface, as compared to other methods, such as, e.~g., de
Haas--van Alphen effect. The resonant absorption is also often
smeared by the dependence of $\omega_c$ on~$p_z$, which is an
additional disadvantage.

We have applied the CR absorption technique to study the cyclotron
motion in the vicinity of the Lifshitz transition in graphite. The
low-temperature in-plane conductivity of this material is
relatively low, $\sigma\sim
10^7-10^8\:(\Omega\cdot\mathrm{m})^{-1}$, and  it quickly
decreases upon the application of a magnetic
field~\cite{Brandt1988,Du2005}. The skin depth thus reaches tens
of nanometers
%well above the limit of graphite.\cite{PartoensPRB06}
and greatly exceeds the spacing between adjacent graphene layers.
Moreover, graphite is a highly anisotropic crystal with rather
flat electronic dispersion in the $z$ direction (perpendicular to
the layers). It appears as a suitable material for CR studies of
the electronic system near the Lifshitz transition driven by the
trigonal warping of electronic bands.

\begin{figure*}
    \begin{minipage}{0.33\linewidth}
      \scalebox{0.9}{\includegraphics{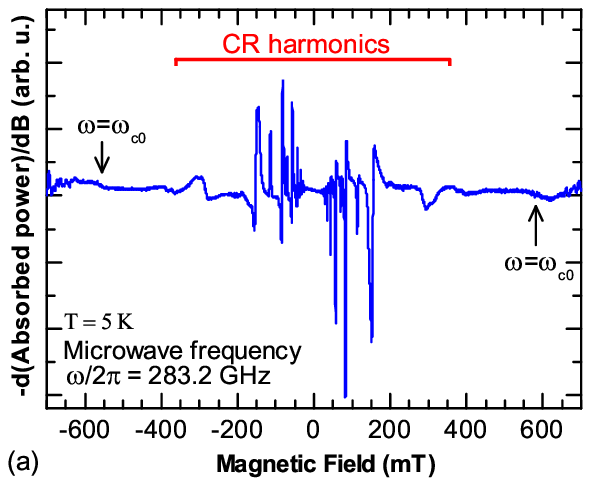}}
    \end{minipage}\hfill
    \begin{minipage}{0.33\linewidth}
      \scalebox{0.27}{\includegraphics{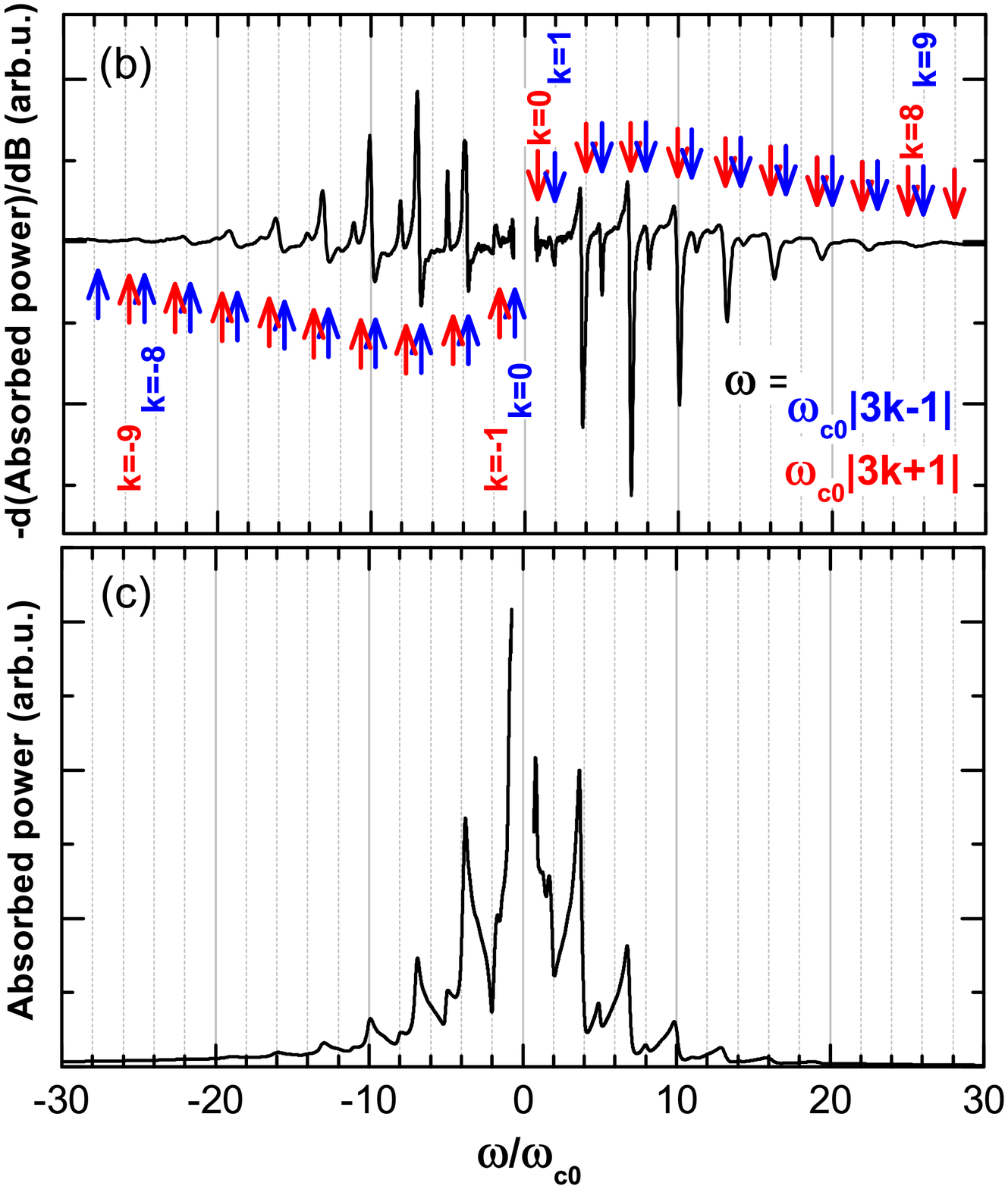}}
    \end{minipage}\hspace{2mm}
     \begin{minipage}{0.32\linewidth}
      \caption{\label{fig:SingleSPKT}
(color online) Magneto-absorption spectra of natural graphite
measured at a fixed microwave excitation energy
$\hbar\omega=1.171$~meV and detected with the help of the field
modulation technique at temperature of 5~K. Harmonics of fundamental CR frequency
$\omega_{c0}=eB/m_{c0}$ ($m_{c0}=0.057m_0$) are observed down to
fields of 20 mT. (a)~Derivative of the absorption with respect to
the magnetic field~$B$, as a function of~$B$. (b)~The same plotted
as a function of $\omega/\omega_{c0}$, so that individual
harmonics at frequencies of $|3k\pm1|\omega_{c0}$,
$k=0,\pm1,\pm2\ldots$, are clearly seen, as marked by vertical
arrows. (c)~Absorption as a function of $\omega/\omega_{c0}$
obtained by the numerical integration of the curve presented in
part (a) with respect to $B$.}
    \end{minipage}
\end{figure*}

CR absorption was measured using the setup routinely applied to
high-frequency electron paramagnetic resonance
experiments~\cite{Neugebauer2010}. A flake of natural graphite
(${50}\,\mu\mbox{m}$ thick, area ${1}\,\mbox{mm}^2$) was placed in
a Fabry-Perot cavity mounted inside a superconducting coil. The
magnetic field was applied perpendicular to the graphene layers.
Linearly polarized microwave radiation from a Gunn diode tripled to frequency of
283.2~GHz (1.171 meV) was delivered to the sample via
quasi-optics waveguides. The field-modulation technique was
applied to enhance the detection sensitivity. The modulation
amplitude was chosen in a way to maximize the signal but to not
distort the measured lineshapes.

A representative experimental spectrum (raw data) is shown in
Fig.~\ref{fig:SingleSPKT}(a). This trace represents the response
of the natural graphite specimen measured as a function of the
magnetic field at fixed microwave frequency. Because of the field
modulation technique, it corresponds to the derivative of the
absorbed power with respect to $B$. The magneto-absorption
response of graphite is expected to be mostly sensitive to
singularities in the electronic joint density of states, located
at the $K$ and $H$ points of the graphite Brillouin zone. A number
of the observed resonances can be easily identified as due to
electronic states at the $K$ point, along the results of previous
similar
studies~\cite{Galt1956,Lax1957,Nozieres1958,Inoue1962,Dresselhaus1974,Doezema1974}.
Holes at the $H$ point as well as decoupled sheets of graphene on
the surface of graphite give rise to resonances at different
spectral range (much lower magnetic
fields)~\cite{OrlitaPRL08,NeugebauerPRL09}.

To a very first approximation, the $K$ point electrons of graphite
have parabolic dispersion. Their effective mass, most frequently
reported to be in the range from $m=0.057m_0$ to
$0.060m_{0}$~\cite{Brandt1988} ($m_0$ free electron mass), fixes the cyclotron frequency at
$\hbar\omega_{c0}\approx{2}\times B[\mbox{T}]\,\mbox{meV}$. Then,
the broad but still visible resonance at $|B|\approx{0}.6$~T is
attributed to the fundamental CR absorption. All other observed
resonances are higher harmonics of the fundamental one. This is
evidenced in Fig.~\ref{fig:SingleSPKT}(b) where the spectrum from
Fig.~\ref{fig:SingleSPKT}(a) is re-plotted against
$\omega/\omega_{c0}$ (i.e. versus $B^{-1}$ instead of $B$).
$\hbar\omega_{c0}$ is eventually set at $2.05\times B$[T] meV. In
agreement with previous reports~\cite{Galt1956}, the observed
harmonics follow two series: $\omega\approx|3k\pm1|\omega_{c0}$,
where $k=0,\pm1,\pm2\ldots$.

The superior quality of the present data (due to higher
frequencies applied and perhaps better quality of graphite
specimens) allows us to uncover more and intriguing spectral
features. Our key observations, that we interpret in the
following, are: i) the appearance of a large number (up to 20) of
CR harmonics ii) an enhanced strength of $3k+1$ harmonics as
compared to the strength of the $3k-1$ series at $B>0$ (and
\textit{vice versa} at $B<0$) and finally, iii) a very
characteristic, asymmetric broadening of the observed resonances,
enhanced on the low frequency (high-field) sides of the absorption
peaks. These features are clearly seen in the raw data and also in
Fig.~\ref{fig:SingleSPKT}(c) in which we reproduce the actual
absorption spectrum, as derived from the numerical integration,
over the magnetic field, of the measured (differential) signal.

The appearance of $n$-harmonics with $n=3k\pm{1}$ is usually
understood as due to breaking of the isotropy of the
electronic spectrum in the layer plane by the trigonal warping. For
isotropic bands, only the $k=0$ fundamental transition
is allowed, whereas the $n=3k\pm{1}$ harmonic appears in the
$|k|$th order of the perturbation theory with respect to the
trigonal warping term~\cite{Inoue1962,Abergel2007}. The spectrum
in Fig.~\ref{fig:SingleSPKT} contains many harmonics which start
to fall off only at large indices $n\gtrsim7$. Clearly, the
perturbation theory is not applicable to interpret these data.
Instead, we will use the quasi-classical approximation.

Eq.~(\ref{Newton=}) can be cast in the Hamiltonian form
in the phase space $(p_x,p_y)$:
%\begin{equation}
${dp_x}/{dt}=-{\partial\mathcal{H}(p_x,p_y)}/{\partial{p}_y}$,
${dp_y}/{dt}={\partial\mathcal{H}(p_x,p_y)}/{\partial{p}_x}$,
%\end{equation}
with the Hamiltonian $\mathcal{H}(p_x,p_y)=-(eB/c)\,\ep(p_x,p_y)$
(we omitted $p_z$, which enters as a parameter). Generally,
classical Hamiltonian systems exhibit a very rich behavior.
However, they share  some universal features when the energy~$\ep$
is close to that of a saddle point of the Hamiltonian,
$\ep=\ep_\mathrm{sp}$, as is well-known in the classical nonlinear
physics~\cite{SagdeevZaslavsky}.
%corresponding
%to a van Hove singularity in the electronic dispersion.
(i)~The cyclotron motion in the vicinity of a saddle point is
slow and its period diverges logarithmically, $\omega_c(\ep)\to{0}$
for $\ep\to\ep_\mathrm{sp}$. (ii)~The
Fourier spectrum of this motion contains many harmonics and their
number diverges when $\ep\to\ep_\mathrm{sp}$.
%and they merge into a continuous spectrum exactly
%at $\ep=\ep_\mathrm{sp}$.
This second fact provides an obvious hint for the interpretation
of the experimental data.
%The first fact also manifests itself in the experimental
%spectrum, although in a less obvious way, as will be seen
%later.

\begin{figure}[t]
%    \begin{minipage}{0.62\linewidth}
      \scalebox{0.45}{\includegraphics{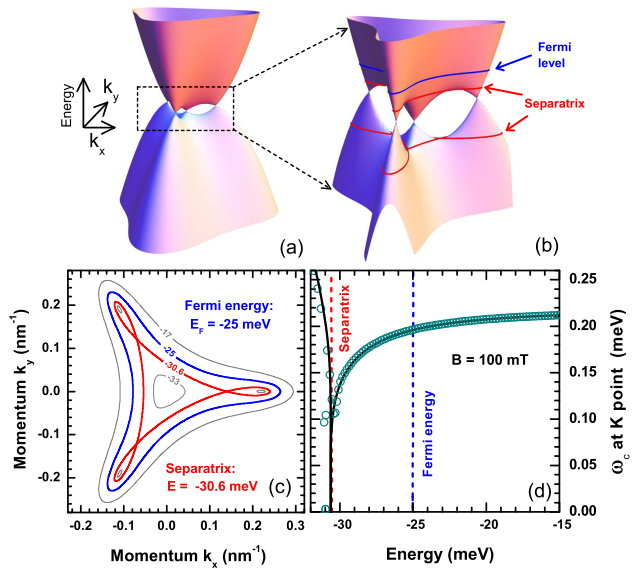}}
%    \end{minipage}\hfill
%    \begin{minipage}{0.32\linewidth}
      \caption{\label{fig:contours}
(color online) (a,b) Electronic structure near the $K$ point of
graphite ($k_z=0$). Two separatrix lines pass through six saddle
points. The Fermi level is located about 6~meV above the upper
separatrix. (c)~Constant energy contours in the $(k_x,k_y)$ plane
for $k_z=0$ for $\ep=-17$, $\ep=-25$ (Fermi level), $\ep=-30.6$
(upper separatrix), and $\ep=-33$~meV. (d)~Classical cyclotron
frequency $\hbar\omega_c(\ep,k_z=0)$ at $B=100\:\mbox{mT}$ in the
relevant energy interval. $\omega_c$ vanishes at the saddle point.
Open circles show the LL spacing, $\Delta\ep_l=\ep_{l+1}-\ep_l$,
as a function of $\ep_l$, derived from the SWM model. Roughly 1
meV away from the saddle point, the circles fall on the classical
curve, $\Delta\ep_l\approx\hbar\omega_c(\ep_l)$.}
%    \end{minipage}
\end{figure}

The experimentally probed electronic states are those around the
Fermi level~$\ep_F$. Thus, the effects discussed above are
important if $\ep_F\approx\ep_\mathrm{sp}$. This is the case of
graphite, as illustrated in Fig.~\ref{fig:contours} using standard
calculations based on the Slonczewski-Weiss-McClure (SWM)
model~\cite{SWM} in the two-band
approximation (see %Appendix~\ref{app:SWM}
Supplementary Information). Here we used the standard values of
the SWM parameters~\cite{Brandt1988}:
$\gamma_0=3150\:\,\mbox{meV}$, $\gamma_1=375\:\,\mbox{meV}$,
$\gamma_2=-20\:\,\mbox{meV}$, $\gamma_3=315\:\,\mbox{meV}$,
$\gamma_4=44\:\,\mbox{meV}$, $\gamma_5=38\:\,\mbox{meV}$,
$\Delta=-8\:\,\mbox{meV}$. The band dispersion has six saddle
points at two different energies $\ep_\mathrm{e-sp}$ and
$\ep_\mathrm{h-sp}$, which define two separatrices -- isoenergetic
lines separating regions with different topology. Fermi level
crossing these saddle points would imply the change in the
topology of the Fermi surface, which actually corresponds to the
Lifshitz transition of the neck-collapsing type. The Fermi level
is close to the upper separatrix, on which we focus our attention
hereafter, $\ep_\mathrm{sp}\equiv\ep_\mathrm{e-sp}$. The single
electron pocket around the $K$ point at $\ep_F>\ep_\mathrm{sp}$,
splits into four disconnected pockets when $\ep_F$ goes below
$\ep_\mathrm{sp}$. Fig.~\ref{fig:contours}(d) shows the classical
cyclotron frequency for the SWM dispersion at $k_z=0$, which
vanishes at $\ep=\ep_\mathrm{sp}$.

In the language of quantum mechanics, the $k_{z}=0$ energy
spectrum consists of discrete Landau levels (LLs) $\ep_l$. In the
quasiclassical approximation, $\ep_l$ can be found from the
Bohr-Sommerfeld quantization rule.
The $n^{th}$ CR harmonic corresponds to the transition over $n$
levels, $n\hbar\omega_c\approx\ep_{l+n}-\ep_l$, to the leading
order in~$\hbar$. The decrease of $\omega_c(\ep)$ at
$\ep\to\ep_\mathrm{sp}$ corresponds to an accumulation of LLs.
Nevertheless, $\Delta\ep$ does not approach zero, since the
condition of the validity of the quasiclassical quantization,
$|\omega_c(\ep+\hbar\omega_c)-\omega_c(\ep)|\ll\omega_c(\ep)$,
holds only if $\ep$ is not too close to $\ep_\mathrm{sp}$. LLs
always remain discrete, see Fig.~\ref{fig:contours}(d).
%The level spacing at
%$\ep\approx\ep_\mathrm{sp}$ can be estimated from
%$\Delta\ep\sim\hbar\omega_c(\ep_\mathrm{sp}+\Delta\ep)$.

As we will show later, $\ep_F-\ep_\mathrm{sp}$ is about 6~meV,
i.e., it is five times larger than the microwave frequency,
$\hbar\omega=1.171\:\mbox{meV}$. Our classical approximation is justified
in this case. As a matter of fact, the quasi-classical approximation works
the better, the smaller $\omega$ is. However, if microwave frequency is too small,
the harmonic structure will be smeared by broadening of electronic states.
The optimal frequency, used in the experiment, is thus determined by
an appropriate compromise between these two competing conditions.

Assuming that the absorbed power is proportional to the real part
of the conductivity, $\Re\sigma_{xx}(\omega)$, and calculating the
latter from the standard kinetic equation~\cite{Abrikosov} in the
simplest relaxation time approximation for the collision integral
(see
%Appendix~\ref{app:kinetic}
Supplementary Information), we obtain
\begin{equation}
\Re\sigma_{xx}(\omega)=\frac{e^2}{\pi^2\hbar}
\sum_{n=-\infty}^\infty\int\left(-\frac{\partial{f}}{\partial\ep}\right)\frac{\Gamma{m}_c|v_{x,n}|^2\,dk_z d\ep}%
{\hbar^2(\omega-n\omega_c)^2+\Gamma^2}
,
\label{abs=}
\end{equation}
where the $k_z$ integration is from $-\pi/(2a_z)$ to $\pi/(2a_z)$.
Both the cyclotron frequency, $\omega_c$, and the cyclotron mass,
$m_c=-eB/(c\omega_c)$, depend on $\ep$ and $k_z$. The basic
frequency $\omega_{c0}$, introduced earlier, is
$\omega_{c0}=\omega_c(\ep=\ep_F,k_z=0)$.
$\vec{v}_n=\vec{v}_n(\ep,k_z)$ is the Fourier harmonic of the
electron velocity, corresponding to the term
$\propto{e}^{-in\omega_ct}$, determined from the
solution of the unperturbed equation of motion,
Eq.~(\ref{Newton=}). Finally, $\Gamma$ accounts for relaxation,
and $f(\ep)$ is the Fermi function.

Even without solving Eq.~(\ref{Newton=}), it is easy to see that
the triangular symmetry of $\ep(\vec{p})$ in the $(p_x,p_y)$ plane
fixes $\vec{v}_n=0$ for $n=3k$, $k=\pm 1,\pm2,\ldots$. In
Fig.~\ref{fig:SingleSPKT}(b), the resonances at $n=3k$ are absent,
which demonstrates that the triangular symmetry is not broken in
graphite. This is in contrast with recent reports for a bilayer
graphene~\cite{Vafek2010,Lemonik2010,Novoselov2011,Mucha-Kruczynski2011},
even though it is formally described by the same single-particle
Hamiltonian (for a fixed $k_z$)~\cite{OrlitaPRL09}. The same
symmetry fixes $v_{n,x}$ to be real and $v_{n,y}=\pm{i}v_{n,x}$
for $n=3k\pm{1}$, so the peaks at $n=3k+1$ and $n=3k-1$ are seen
in the opposite circular polarizations of the microwave field.
This helps us to understand the observed difference in the
intensities of the $n=3k+1$ and $n=3k-1$ series. Indeed, for
$B>0$, when the electron moves along the Fermi surface, shown in
Fig.~\ref{fig:contours}(b), in the overall counterclockwise
direction, it should be more strongly coupled to the
counterclockwise polarized radiation. As both circular
polarizations are equally present in the incoming linearly
polarized radiation, the spectrum is fairly symmetric with respect
to $B\to-B$.

The present studies are restricted to bulk graphite, a system with fixed
Fermi level but in an apparent proximity to the Lifshitz transition.
An obvious experimental challenge would be to trace the CR response when
changing the Fermi level with respect to the separatrix energy, with an
attempt to tune the proximity to Lifshitz transition in graphitic structures.
This can be in principle envisaged for electrostatically gated bilayer
graphene~\cite{Mucha-Kruczynski2011,Novoselov2011} and/or for bulk graphite under hydrostatic pressure~\cite{MendezPRB80}.
Importantly, such experiments require no degradation of the
quality of the sample, which likely excludes the experiments on, for example, chemically doped structures.

Besides the large number of harmonics, typical of a classical
motion near a saddle point, the proximity to the Lifshitz
transition also leads to some lowering of the cyclotron frequency.
Indeed, the fundamental cyclotron frequency determined from
the period in $B^{-1}$ of the spectrum,
$\hbar\omega_{c0}/B=2.05\:\mbox{meV/T}$, is slightly lower than
its parabolic-band limit at $k_z=0$, $2.24\:\mbox{meV/T}$. The
latter value, however, relies on the specific values of the
parameters of the SWM model. More apparent effects are deduced
from the analysis of the peak shapes (which are determined by the
integration over~$k_z$).

\begin{figure}
\scalebox{0.42}{\includegraphics{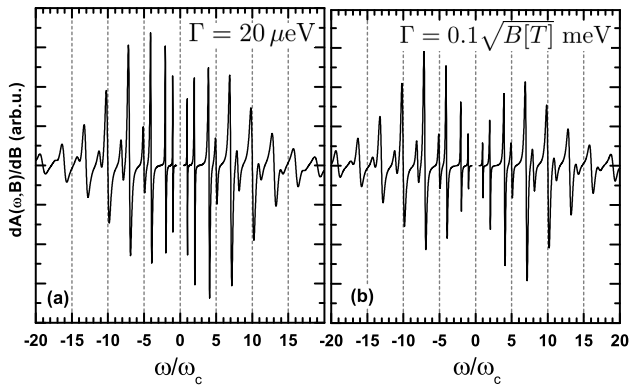}}
\caption{\label{fig:Theory} Derivative of the absorption with
respect to~$B$, as calculated using Eq.~(\ref{abs=}) for different
electronic broadenings~$\Gamma$. (a)~a constant broadening
$\Gamma=20\:\mu\mbox{eV}$ and
(b)~$\Gamma=0.1\sqrt{B[\mathrm{T}]}\:\mbox{meV}$.}
\end{figure}

As seen from Fig.~\ref{fig:SingleSPKT}(c), each peak has an abrupt
cut-off on the \emph{high-frequency} side and a tail on the
\emph{low-frequency} side. This contradicts the first intuition,
based on the well-known fact that the parabolic part of the bands
becomes steeper as $k_z$ increases from the $K$~point towards the
$H$~point. This pushes the LLs upwards as $k_z$ increases and
would result in a tail on the \emph{high-frequency} side of each
peak in the absorption spectrum $\mathcal{A}(\omega,B)$~\cite{Faugeras2012}. However, the
bottom of the conduction band [defined as $\ep(\vec{p}=0,k_z)$]
and the saddle point $\ep_\mathrm{sp}(k_z)$ shift upwards upon
increasing $k_z$, as $\ep(\vec{p}=0,k_z)=2\gamma_2\cos{k}_za_z$,
$\gamma_2<0$. Thus, the Fermi level approaches the saddle point as
$k_z$ moves away from $k_{z}=0$ point and $\omega_c(\ep_F,k_z)$
decreases simultaneously, see Fig.~\ref{fig:contours}(d). This
provides a tail on the \emph{low-frequency} side of the peaks.
Thus, the suppression of $\omega_c$ near the Lifshitz transition
is crucial to interpret the peak asymmetry.

The spectrum derived from Eq.~(\ref{abs=}) is shown in
Fig.~\ref{fig:Theory}, with $\ep_F$ and $\Gamma$ as the only
adjustable parameters -- the parameters of the SWM model were
fixed~\cite{Brandt1988}. The best agreement is obtained for
$\ep_F=-25\:\mbox{meV}$. If $\ep_F=-24\:\mbox{meV}$, the peaks
have no asymmetry, since $\omega_c(\ep_F=-24\,\mbox{meV},k_z)$ has
a significant upturn on increasing~$k_z$. When
$\ep_F=-26\,\mbox{meV}$ the fall-off of large-$n$ harmonics is
noticeably slower than the experimental one. In other words, the
closer $\ep_F$ is to $\ep_\mathrm{sp}$, the more harmonics are
seen in the spectrum. The frequency
$\omega_c(\ep_F=-25\,\mbox{meV},k_z=0)/B=2.03\,\mbox{meV/T}$
agrees with the experimental value, $2.05\:\mbox{meV/T}$. The
value $\ep_F=-25\,\mbox{meV}$ is also in good agreement with the
one determined independently from the charge neutrality condition
(see Supplementary Information), $\ep_F=-24\,\mbox{meV}$. A
constant value of $\Gamma=20\,\mu\mbox{eV}$ was assumed for the
curve in Fig.~\ref{fig:Theory}(a). Apparently, this does not
describe well the amplitudes of peaks at $n=1,2$: the theoretical
peaks are narrower and thus more intense than the experimental
ones. Better agreement is obtained under assumption that
$\Gamma\propto\sqrt{B}$ (see Ref.~\onlinecite{AndoJPSJ1975} and
Supplementary Information). Notably, the curve in
Fig.~\ref{fig:Theory}(b) with
$\Gamma=0.1\sqrt{B[\mathrm{T}]}\:\mbox{meV}$ corresponds to the
zero-field relaxation rate $\hbar/\tau_{B=0}=40\:\mu\mbox{eV}$.
The extracted value of  $\tau_{B=0}$  provides the zero-field dc
conductivity  $\sigma=1.2\times10^8\:(\Omega\cdot\mbox{m})^{-1}$ (see Supplementary Information).
This is fully consistent with typical literature data~\cite{Du2005,Brandt1988}
and implies a mean electron free path of 6~$\mu$m, which is, notably,
comparable or even longer  than the corresponding values reported
for strictly 2D graphene-based structures~\cite{Novoselov2011,BolotinPRL08,MayorovNL11}.

To conclude, we have introduced CR experiments as a new tool to
study Lifshitz transitions. We have shown how the proximity to the
Lifshitz transition manifests itself in the CR spectrum of a model
system, bulk graphite. Namely, we have observed a multi-mode
response, where the basic CR mode is accompanied by many
harmonics. Using the standard SWM model for the electronic band
structure of graphite to analyze the data, we have determined the
Fermi energy and estimated the electronic broadening. The
similarity between the band structure of graphite near the
$K$~point and that of a bilayer graphene logically suggests to
probe Lifshitz transition in the latter system by CR methods, and
to shed more light on the currently debated issue of spontaneous symmetry
breaking in bilayer
graphene~\cite{Vafek2010,Lemonik2010,Novoselov2011,Mucha-Kruczynski2011}.

We thank M.-O. Goerbig and J.-N. Fuchs for collaboration on the
early stages of this work, and V. F. Gantmakher and Yu. I.
Latyshev for helpful discussions. Part of this work was supported
by RTRA ``DISPOGRAPH'' project. M.O. acknowledges support from
GACR P204/10/1020 and GRA/10/E006 (EPIGRAT). F.~M.~D.~P. thanks
LPMMC for hospitality.

\begin{widetext}
\begin{center}
{\large
\textbf{Supplementary Information for}\\\vspace{2mm}
\emph{Cyclotron motion in the vicinity of Lifshitz transition in graphite}\\\vspace{2mm}
by M. Orlita, P. Neugebauer, C. Faugeras, A.-L. Barra, M. Potemski, F. M. D. Pellegrino,
and D. M. Basko}
\end{center}
\end{widetext}

\section{Slonczewski-Weiss-McClure model and its two-band projection}
\label{app:SWM}

The Hamiltonian of the $AB$-stacked graphite can be written as
\begin{equation}
H=
\left(\begin{array}{cccccccccc}
\ldots & \ldots & \ldots & \ldots & \ldots & \ldots & \ldots & \ldots
& \ldots & \ldots \\
\ldots & V_{11} & V_{12}^\dagger & H_1 & V_{12}^\dagger
& V_{11}^\dagger & 0 & 0 & 0 & \ldots \\
\ldots & 0 & V_{22} & V_{12} & H_2 & V_{12} & V_{22}^\dagger & 0 & 0 & \ldots \\
\ldots & 0 & 0 & V_{11} & V_{12}^\dagger & H_1 & V_{12}^\dagger &
V_{11}^\dagger & 0 & \ldots \\
\ldots & 0 & 0 & 0 & V_{22} & V_{12} & H_2 & V_{12} & V_{22}^\dagger & \ldots \\
\ldots & \ldots & \ldots & \ldots & \ldots & \ldots & \ldots & \ldots
& \ldots & \ldots
\end{array}\right),
\end{equation}
where each element is a matrix in the Hilbert space of a
single layer.
$H_1$ and $H_2$ contain the in-plane nearest-neighbor
coupling with the matrix element~$\gamma_0$ as well as
the diagonal shift~$\Delta$.
$V_{12}$ represents the coupling between neighboring
layers with matrix elements $\gamma_1$ for the neighboring
atoms, and $\gamma_3,\gamma_4$ for the second nearest
neighbors.
$V_{11}$ and $V_{22}$ correspond to second-nearest-layer
coupling with matrix elements $\gamma_2/2$ and $\gamma_5/2$.
Due to the translational invariance in the $z$ direction,
the problem can be reduced to that of an effective bilayer:
\begin{widetext}
\begin{equation}
H_{k_z}=
\left[\begin{array}{cc}
H_1+V_{11}e^{2ik_za_z}+V_{11}^\dagger{e}^{-2ik_za_z} &
2V_{12}^\dagger\cos{k}_za_z \\
2V_{12}\cos{k}_za_z &
H_2+V_{22}e^{2ik_za_z}+V_{22}^\dagger{e}^{-2ik_za_z}
\end{array}\right],
\end{equation}
\end{widetext}
where $a_z$ is the distance between the neighboring layers,
and $k_z$ is the wave vector in the direction perpendicular
to the layers, $-\pi/(2a_z)<{k}_z\leq\pi/(2a_z)$ (note that
the period of the structure is $2a_z$, that is, two layers).
Expansion of the single-layer Hamiltonian in $p_x,p_y$, the
in-plane quasi-momentum components counted from the $H-K-H$
line, gives
\begin{equation}
{H}_{k_z}(\hat{\vec{p}})=\left[\begin{array}{cccc}
\Gamma_2 & v\hat{p}_- &
-\alpha_4v\hat{p}_- &
\alpha_3v\hat{p}_+ \\
v\hat{p}_+ & \Gamma_5 & \Gamma_1 &
-\alpha_4v\hat{p}_- \\
-\alpha_4v\hat{p}_+ &
\Gamma_1 & \Gamma_5 & v\hat{p}_-\\
\alpha_3v\hat{p}_- &
-\alpha_4v\hat{p}_+ &
v\hat{p}_+ & \Gamma_2 \end{array}\right],
\end{equation}
where
\begin{eqnarray*}
&&v=\frac{3}{2}\,\frac{\gamma_0{a}}{\hbar},\quad
\Gamma_1=2\gamma_1\mathcal{C},\quad
\Gamma_2=2\gamma_2\mathcal{C}^2,\\
&&
\alpha_{3,4}=\frac{2\gamma_{3,4}}{\gamma_0}\,\mathcal{C},\quad
\Gamma_5=2\gamma_5\mathcal{C}^2+\Delta,\quad
\mathcal{C}\equiv\cos{k}_za_z,
\end{eqnarray*}
$a=1.42\:\mbox{\AA}$ is the distance between the
neighboring carbon atoms in the same layer,
and $\hat{p}_\pm=-i\hbar(\partial_x\pm{i}\partial_y)$.
It is convenient to rotate the basis in the space of the
4-columns $(\psi_1,\psi_2,\psi_3,\psi_4)^T$ as
\begin{equation}
\psi_1'=\frac{\psi_2+\psi_3}{\sqrt{2}},\quad
\psi_2'=\psi_1,\quad\psi_3'=\psi_4,\quad
\psi_4'=\frac{\psi_2-\psi_3}{\sqrt{2}},
\end{equation}
so the transformed Hamiltonian becomes
\begin{equation}
H'_{k_z}(\hat{\vec{p}})=
\left[\begin{array}{cccc}
\Gamma_5+\Gamma_1 & \bar{v}_4\hat{p}_+ &
\bar{v}_4\hat{p}_- & 0 \\
\bar{v}_4\hat{p}_- & \Gamma_2 & \alpha_3v\hat{p}_+ &
v_4\hat{p}_- \\
\bar{v}_4\hat{p}_+ &
\alpha_3v\hat{p}_- & \Gamma_2 & -v_4\hat{p}_+\\
0 & v_4\hat{p}_+ &
-v_4\hat{p}_- & \Gamma_5-\Gamma_1 \end{array}\right],
\end{equation}
where $v_4=v(1+\alpha_4)/\sqrt{2}$, $\bar{v}_4=v(1-\alpha_4)/\sqrt{2}$.
The four eigenvalues for each $p_x,p_y,k_z$ can be found by calculating
the determinant
\begin{equation}\begin{split}
&\det{H}'_{k_z}(\vec{p})=(1-\alpha_4^2)^2(vp)^4-\\
&{}\quad-2\alpha_3\left[(1+\alpha_4^2)\Gamma_1+2\alpha_4\Gamma_5\right]
(vp)^3\cos{3}\varphi_\vec{p}+\\
&{}\quad+\left[\alpha_3^2(\Gamma_1^2-\Gamma_5^2)-2(1+\alpha_4^2)\Gamma_2\Gamma_5
-4\alpha_4\Gamma_2\Gamma_1\right](vp)^2-\\
&{}\quad-\Gamma_2^2(\Gamma_1^2-\Gamma_5^2),
\end{split}\end{equation}
and replacing $\Gamma_5\to\Gamma_5-\ep$,
$\Gamma_2\to\Gamma_2-\ep$ to obtain $\det(H'_{k_z}-\ep)$.

\begin{figure}
\vspace*{5mm}
\includegraphics[width=8cm]{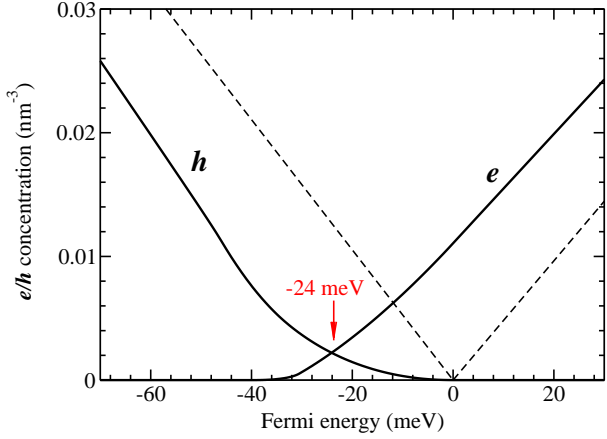}
\caption{\label{fig:densities}
Electron and hole concentrations as functions of the Fermi energy,
as given by Eqs.~(\ref{ne=}), (\ref{nh=}) (thick solid lines),
together with the simplified expression~(\ref{neh=}) (dashed lines).
}
\end{figure}

To obtain the effective $2\times{2}$ Hamiltonian, acting
in the subspace of the two low-energy bands, we eliminate
$\psi_1',\psi_4'$ from the Schr\"odinger equation
$H_{k_z}'(\hat{\vec{p}})\,\psi'=\ep\psi'$. This gives
\begin{equation}\begin{split}
\hat{H}^{2\times{2}}_{k_z}(\hat{\vec{p}})={}&
\left[\begin{array}{cc}
\Gamma_2 & \alpha_3v\hat{p}_+-v^2\hat{p}_-^2/\Gamma_1 \\
\alpha_3v\hat{p}_--v^2\hat{p}_+^2/\Gamma_1 & \Gamma_2
\end{array}\right]-\nonumber\\
{}&+\frac{2\zeta_{eh}}{\Gamma_1}
\left[\begin{array}{cc}
v^2\hat{p}_-\hat{p}_+ & 0 \\ 0 & v^2\hat{p}_+\hat{p}_-
\end{array}\right],
\end{split}\end{equation}
where
\begin{equation}
\zeta_{eh}=\alpha_4+\frac{\Gamma_5-\Gamma_2}{2\Gamma_1}
\end{equation}
determines the electron-hole asymmetry of the spectrum.
At energies we are interested in (a few tens of meV)
the two-band approximation works well for all $k_z$ except
the immediate vicinity of the $H$ point,
$\pi/2-|k_z|a_z\lesssim{0}.05$. In the two-band approximation,
there is an analytical expression for the energies, which is
most conveniently written in the polar coordinates,
$p_x=p\cos\varphi$, $p_y=p\sin\varphi$:
\begin{equation}\begin{split}\label{eptwoband=}
&\ep_{k_z}^\pm(p\cos\varphi,p\sin\varphi)=
\Gamma_2+2\zeta_{eh}\,\frac{(vp)^2}{\Gamma_1}\pm\\
&\qquad\pm\sqrt{\frac{(vp)^4}{\Gamma_1^2}
-2\alpha_3\frac{(vp)^3}{\Gamma_1}\cos{3\varphi}
+\alpha_3^2(vp)^2}.
\end{split}\end{equation}
From this expression one can deduce the energies of the
saddle point in the conduction band, the energies of the
three ``leg'' conical points, where the conduction and the
valence bands touch each other (in addition to the conical
point at $\vec{p}=0$ with $\ep=\Gamma_2$), and the energy
of the saddle point in the valence band:
\begin{subequations}\begin{eqnarray}
&&\ep_{\mathrm{e-sp}}=\Gamma_2+\frac{\alpha_3^2\Gamma_1}{4(1-2\zeta_{eh})},\\
&&\ep_\mathrm{leg}=\Gamma_2+2\zeta_{eh}\alpha_3^2\Gamma_1,\\
&&\ep_{\mathrm{h-sp}}=\ep_c-\frac{\alpha_3^2\Gamma_1}{4(1+2\zeta_{eh})}.
\end{eqnarray}\end{subequations}
Given the dispersion~(\ref{eptwoband=}), one can determine
the concentrations of electrons and holes at zero temperature as
functions of the Fermi energy,
\begin{subequations}
\begin{eqnarray}
&&n_e(\ep_F)=4
\int\limits_{\ep^+(\vec{p})<\ep_F}\frac{d^3\vec{p}}{(2\pi\hbar)^3},
\label{ne=}\\
&&n_h(\ep_F)=4
\int\limits_{\ep^-(\vec{p})>\ep_F}\frac{d^3\vec{p}}{(2\pi\hbar)^3},
\label{nh=}
\end{eqnarray}
\end{subequations}
where the factor of 4 takes care of valley and spin degeneracy.
If the sample is undoped, $\ep_F$ can be determined from the
neutrality condition $n_e(\ep_F)=n_h(\ep_F)$. Electron and
hole concentrations obtained from Eqs.~(\ref{ne=}), (\ref{nh=}),
are plotted in Fig.~\ref{fig:densities}, together with the
simplified expression,
\begin{equation}\begin{split}
&n_{e,h}(\ep_F)=\frac{2\gamma_1|\ep_F|\,\theta(\pm\ep_F))}%
{\pi^2a_z(3\gamma_0a/2)^2}\,\mathcal{Z}(\pm{4}\gamma_4/\gamma_0),\\
&\mathcal{Z}(x)\equiv\frac{2}x\left(\frac{\pi}{4}
-\frac{1}{\sqrt{1-x^2}}\arctan\sqrt{\frac{1-x}{1+x}}\right),
\label{neh=}
\end{split}\end{equation}
obtained by neglecting $\gamma_2,\gamma_3,\gamma_5$.

\section{Conductivity from the kinetic equation}
\label{app:kinetic}

We determine the conductivity from the electronic dispersion
following the standard procedure\cite{Abrikosov}.
First, let us consider the problem at some fixed value of~$k_z$,
and find the in-plane electric current produced by an external
in-plane electric field
\begin{equation}\label{Et=}
\vec{E}(t)=\vec{E}_\omega\,e^{-i\omega{t}}
+\vec{E}_\omega^*\,e^{i\omega{t}}.
\end{equation}
The in-plane group velocity of the electrons being
$\vec{v}=\partial\ep(\vec{p})/\partial{p}$ ($\vec{p}$ is in the
$xy$ plane, and we omit the $k_z$ argument for the moment),
the semiclassical equations of motion are
\begin{subequations}
\begin{eqnarray}
&&\frac{dp_x}{dt}=\frac{eB}c\frac{\partial\ep(p_x,p_y)}{\partial{p}_y}
+eE_x(t),\\
&&\frac{dp_y}{dt}=-\frac{eB}c\frac{\partial\ep(p_x,p_y)}{\partial{p}_x}
+eE_y(t).
\end{eqnarray}
\end{subequations}
Here the electron charge $e<0$.
These equations can be written in the Hamiltonian form
$dp_x/dt=-\partial\mathcal{H}/\partial{p}_y$,
$dp_y/dt=\partial\mathcal{H}/\partial{p}_x$,
with the Hamiltonian function
\begin{equation}\label{Hpxpy=}
\mathcal{H}(p_x,p_y)=-\frac{eB}c\,\ep(p_x,p_y)
+e(E_yp_x-E_xp_y).
\end{equation}
In the absence of the perturbing electric field the
trajectories in the $(p_x,p_y)$-space (cyclotron orbits)
coincide with the constant energy contours,
$\ep(\vec{p})=\mathrm{const}$.
The solution of the unperturbed problem ($\vec{E}=0$)
can be written as
%\begin{subequations}\begin{eqnarray}
%&&p_x(t)=\sum_{n=-\infty}^\infty{P}_n(E)e^{-in\omega_c(E)\,t}
%&&p_y(t)=\sum_{n=-\infty}^\infty{Q}_n(E)e^{-in\omega_c(E)\,t}
%\end{eqnarray}\end{subequations}
\begin{equation}\label{p=pn}
\vec{p}(t)=\sum_{n=-\infty}^\infty
\vec{p}_n(\ep)\,e^{-in\omega_c(\ep)t},\quad
\vec{p}_n(\ep)=\vec{p}_{-n}^*(\ep),
\end{equation}
where the frequency is determined by the derivative of the area,
enclosed by the cyclotron orbit, with respect to energy:
%\begin{equation}
%\frac{2\pi}{\omega_c(\ep)}=-\frac{c}{eB}\frac{\partial}{\partial\ep}
%\int\limits_{\ep(\vec{p})<\ep}dp_x\,dp_y.
%\end{equation}
\begin{equation}
\omega_c(\ep)=-\frac{eB}{m_c(\ep)c},\quad
m_c(\ep)=\frac{1}{2\pi}\frac{\partial}{\partial\ep}
\int\limits_{\ep(\vec{p})<\ep}dp_x\,dp_y.
\end{equation}
Thus defined $m_c$ is nothing but the cyclotron mass.
The harmonics of the velocity can be related to those
of momentum directly from the equations of motion,
which gives
\begin{equation}
v_{nx}(\ep)=-\frac{in}{m_c(\ep)}\,p_{ny}(\ep),\quad
v_{ny}(\ep)=\frac{in}{m_c(\ep)}\,p_{nx}(\ep).
\end{equation}
The symmetry of the cyclotron orbit with respect to
rotations by $2\pi/3$ fixes
\begin{equation}\label{p3k=0}
p_{3k,x}=p_{3k,y}=0,\quad
p_{3k\pm{1},x}=\mp{i}p_{3k\pm{1},y}=P_{3k\pm{1}},
\end{equation}
for integer~$k$, where $P_n(\ep)$ are real.
When $\ep_\mathrm{h-sp}<\ep<\ep_\mathrm{e-sp}$, in addition to the
orbit which is $C_{3v}$-symmetric around the origin, for each~$\ep$
there are three other orbits encircling the three ``leg'' conical
points, each such orbit having only one mirror reflection symmetry.
These give rise to another series solutions with
\begin{equation}
p_{n,x}=P_n',\quad p_{n,y}=iQ_n',
\end{equation}
where $P_n'(\ep)$, $Q_n'(\ep)$ are also real. However, we will
be mostly concerned with the first solution, Eq.~(\ref{p3k=0}).

\begin{figure}
\vspace*{5mm}
\includegraphics[width=8cm]{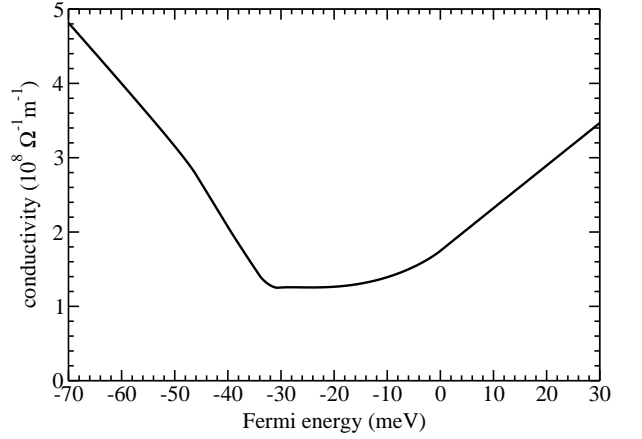}
\caption{\label{fig:conductivity}
The in-plane dc conductivity of graphite at zero temperature
and zero magnetic field for $\hbar/\tau=40\:\mu\mbox{eV}$,
as a function of the Fermi energy.
}
\end{figure}

In order to separate the oscillating motion, it is convenient
to perform the canonical change to action-angle variables of
the unperturbed Hamiltonian, $(p_x,p_y)\to(S,\phi)$. The action
variable is defined by
\begin{equation}\label{Sep=}
S(\ep)=-\frac{c}{eB}
\int\limits_{\ep(\vec{p})<\ep}\frac{dp_x\,dp_y}{2\pi},
\end{equation}
and the angle $\phi=\omega_ct$ for the unperturbed motion.
The inverse transformation, $(S,\phi)\to(p_x,p_y)$, is given
by Eq.~(\ref{p=pn}) with the replacement $\omega_ct\to\phi$
and $\ep\to\ep(S)$, the latter determined from Eq.~(\ref{Sep=}).
As for any canonical transformation, the phase volume is
preserved: $dp_x\,dp_y=dS\,d\phi$.
In the new variables, the Hamiltonian~(\ref{Hpxpy=}) assumes
the form
\begin{equation}
\mathcal{H}(S,\phi)=\int\limits^S\omega_c(S')\,dS'
+\sum_{n\neq{0}}\frac{m_c(S)}{in}\,e\vec{E}\cdot\vec{v}_n(S)\,e^{-in\phi}.
\end{equation}
The distribution function $\mathcal{F}(S,\phi)$ is defined
as the average number of particles in the quantum of the
phase volume, $(2\pi\hbar)^2$, around some given values of
$S,\phi$. It determines the total current carried by the
electrons:
\begin{equation}\label{j=F}
\vec{j}(t)=\int\frac{dS\,d\phi}{(2\pi\hbar)^2}\,
\mathcal{F}(S,\phi,t)\sum_{n=-\infty}^\infty
e\vec{v}_n(S)e^{-in\phi}.
\end{equation}
The distribution function $\mathcal{F}$ satisfies the
kinetic equation
\begin{equation}
\frac{\partial\mathcal{F}}{\partial{t}}+
\frac{\partial\mathcal{H}}{\partial{S}}
\frac{\partial\mathcal{F}}{\partial\phi}-
\frac{\partial\mathcal{H}}{\partial\phi}
\frac{\partial\mathcal{F}}{\partial{S}}=\mathrm{St}[\mathcal{F}],
\end{equation}
where the left-hand side is the canonical Liouville operator,
and the right-hand side represents the collision integral.
We seek the distribution function in the form
$\mathcal{F}(S,\phi,t)=\mathcal{F}_0(S)+\delta\mathcal{F}(S,\phi,t)$,
where $\mathcal{F}_0(S)$ is the equilibrium distribution
function, corresponding to the Fermi-Dirac distribution $f(\ep)$:
\begin{equation}
\mathcal{F}_0(S)=f(\ep(S))
=\frac{1}{1+e^{[\ep(S)-\ep_F]/T}},
\end{equation}
and the correction $\delta\mathcal{F}$ is sought to linear
order in $\vec{E}$.
For the collision term we adopt the simplest approximation
with relaxation time~$\tau$, so the linearized kinetic equation
for $\delta\mathcal{F}$ has the form:
\begin{equation}
\frac{\partial\delta\mathcal{F}}{\partial{t}}
+\omega_c(S)\,\frac{\partial\delta\mathcal{F}}{\partial\phi}
+\frac{\partial\mathcal{F}_0}{\partial{S}}
\sum_nem_c\vec{E}\cdot\vec{v}_ne^{-in\phi}
=-\frac{\delta\mathcal{F}}\tau,
\end{equation}
For the monochromatic field, Eq.~(\ref{Et=}), one readily
finds the oscillating correction
$\delta\mathcal{F}(t)=\delta\mathcal{F}_\omega{e}^{-i\omega{t}}
+\delta\mathcal{F}_\omega^*{e}^{i\omega{t}}$, with
$\delta\mathcal{F}_\omega$ given by
\begin{equation}
\delta\mathcal{F}_\omega=-\frac{\partial\mathcal{F}_0}{\partial{S}}
\sum_{n=-\infty}^\infty
\frac{em_c\vec{v}_n\cdot\vec{E}_\omega}{1/\tau-i(\omega+n\omega_c)}\,
e^{-in\phi}.
\end{equation}
Substituting this expression in Eq.~(\ref{j=F}) and passing from
integration over~$S$ to integration over~$\ep$, we obtain the
current,
$\vec{j}(t)=\vec{j}_\omega{e}^{-i\omega{t}}
+\vec{j}_\omega^*{e}^{i\omega{t}}$, with $\vec{j}_\omega$
given by
\begin{equation}
\vec{j}_\omega
=\frac{e^2}{2\pi\hbar^2}\int{d\ep}
\left(-\frac{\partial{f}}{\partial\ep}\right)
\sum_{n=-\infty}^\infty
\frac{m_c\vec{v}_n(\vec{v}_n^*\cdot\vec{E}_\omega)}{1/\tau-i(\omega-n\omega_c)}.
\end{equation}
Up to now we considered the problem at fixed $k_z$ (recall that
$m_c,\omega_c,\vec{v}_n$ depend on~$k_z$ via $\cos{k}_za_z$).
The obtained current should be summed over~$k_z$ with the help of
\begin{equation}
\sum_{k_z}\to\int\limits_{-\pi/(2a_z)}^{\pi/(2a_z)}\frac{N_za_z\,dk_z}{2\pi},
\end{equation}
where $N_z$ is the number of graphene layers, and thus $N_za_z$
is the thickness of the sample. Also, the result should be
multiplied by the spin multiplicity~2, and by the valley
multiplicity $\mathcal{N}=2$. This gives the in-plane conductivity
per graphene layer, which should be further divided by~$a_z$ to
obtain the bulk conductivity,
\begin{equation}\begin{split}
\sigma_{ij}(\omega)={}&\frac{2\mathcal{N}e^2}{2\pi\hbar^2a_z}
\int{d}\ep
\left(-\frac{\partial{f}}{\partial\ep}\right)
\int\limits_{\pi/2}^{\pi/2}\frac{d(k_za_z)}{2\pi}\times{}\\
{}&\times\sum_{n=-\infty}^\infty\frac{{m}_cv_{i,n}(v_{j,n})^*}%
{\Gamma-i(\omega-n\omega_c)},
\label{sigmacl=}
\end{split}\end{equation}
where $\Gamma=1/\tau$.
The $B\to{0}$ limit is recovered by setting $\omega_c\to{0}$
and
\begin{equation}
\sum_{n=-\infty}^\infty{v}_{i,n}v_{j,n}^*\to
\langle{v}_iv_j\rangle_\ep\equiv
\frac{\int{v}_iv_j\delta(\ep_{k_z}(\vec{p})-\ep)\,dp_x\,dp_y}%
{\int\delta(\ep_{k_z}(\vec{p})-\ep)\,dp_x\,dp_y}.
\end{equation}
We evaluate this integral numerically, and plot the corresponding
dc conductivity in zero magnetic field in Fig.~\ref{fig:conductivity}.

\section{Self-consistent Born approximation in high magnetic fields
for arbitrary electronic dispersion}
\label{app:SCBA}

Here we show that the self-consistent Born approximation, studied
in detail for two-dimensional electron gas with parabolic dispersion
by Ando and Uemura\cite{Ando1,AndoJPSJ1975}, can be straightforwardly
generalized to arbitrary electronic dispersion for short-range
disorder and strong magnetic field. Moreover, the expressions obtained
by Ando and Uemura for the broadening of the Landau levels and the
cyclotron resonance lines remain unchanged.

Consider two-dimensional electrons with a dispersion law $\ep(\vec{p})$.
The uniform magnetic field $\vec{B}$ is perpendicular to the plane, and
is described by the vector potential~$\vec{A}$. In the presence
of a static disordered potential $V(\vec{r})$, the electronic Hamiltonian
takes the form
\begin{equation}
\hat{H}=\ep(-i\hbar\vec\nabla-e\vec{A}/c)+V(\vec{r})\
\equiv\hat{{H}}_0+\hat{{H}}_{dis}.
\end{equation}
The random potential is assumed to be Gaussian with zero average,
$\overline{V(\vec{r})}=0$, and is determined by its pair correlator,
\begin{equation}\label{VV=}
\overline{V(\vec{r})\,V(\vec{r}')}=W(\vec{r}-\vec{r}')
=W_0\,\delta(\vec{r}-\vec{r}'),
\end{equation}
where the overbar denotes the average over the disorder realizations.
If the disorder is due to random impurities with the two-dimensional
density $n_\mathrm{imp}$, and each impurity has a short-range potential
$u_\mathrm{imp}(\vec{r})$, then
\begin{equation}
W_0=n_\mathrm{imp}\left|\int{u}_\mathrm{imp}(\vec{r})\,d^2\vec{r}\right|^2.
\end{equation}

Let us define the disorder-free Green's function:
\begin{equation}\begin{split}
{G}_0(\vec{r},\vec{r}';\ep)&=
\langle\vec{r}|(\ep-\hat{{H}}_0)^{-1}|\vec{r}'\rangle=\\
&=\sum_{l,\alpha}
\frac{\psi_{l\alpha}(\vec{r})\,\psi_{l\alpha}^*(\vec{r}')}{\ep-\ep_l}
\equiv\sum_l\frac{\mathcal{P}_l(\vec{r},\vec{r}')}{\ep-\ep_l}.
\end{split}\end{equation}
Here $l$~labels the Landau levels with energies $\ep_l$, and $\alpha$
labels states on the same Landau level with wave functions
$\psi_{l\alpha}(\vec{r})$ (eigenfunctions of $\hat{{H}}_0$).
For example, in the Landau gauge, $A_x=-By$, they can be labeled by
the momentum~$p_x$:
\begin{subequations}\begin{equation}\label{psi=chi}
\psi_{lp_x}(x,y)=e^{ip_x/\hbar}\chi_l(y+cp_x/eB),
\end{equation}
where the functions $\chi_l(y)$ and Landau level energies $\ep_l$
are found from the following equation:
\begin{equation}
\ep(eBy/c,-i\hbar\partial_y)\,\chi_l(y)=\ep_l\chi(y).
\end{equation}\end{subequations}
The choice of basis in the degenerate manifold of each Landau
level, as well as the specific form of the wave functions will
not be important. We will only need the following general
properties:
(i)~the projector kernel $\mathcal{P}_l(\vec{r},\vec{r}')$
depends only on the relative coordinate $\vec{r}-\vec{r}'$
(and so does the Green's function~${G}_0$), and
(ii)~the degeneracy of each Landau level is
$L_xL_y/(2\pi\ell_B^2)$, where $L_x$ and $L_y$ are the dimensions
of the sample in the $x$ and $y$ directions, respectively, and
$\ell_B^2=\hbar{c}/(eB)$. Thus,
\begin{equation}\label{Pnrr=}
\mathcal{P}_l(\vec{r},\vec{r})=\frac{1}{2\pi\ell_B^2}.
\end{equation}

\begin{figure}
\includegraphics[width=8cm]{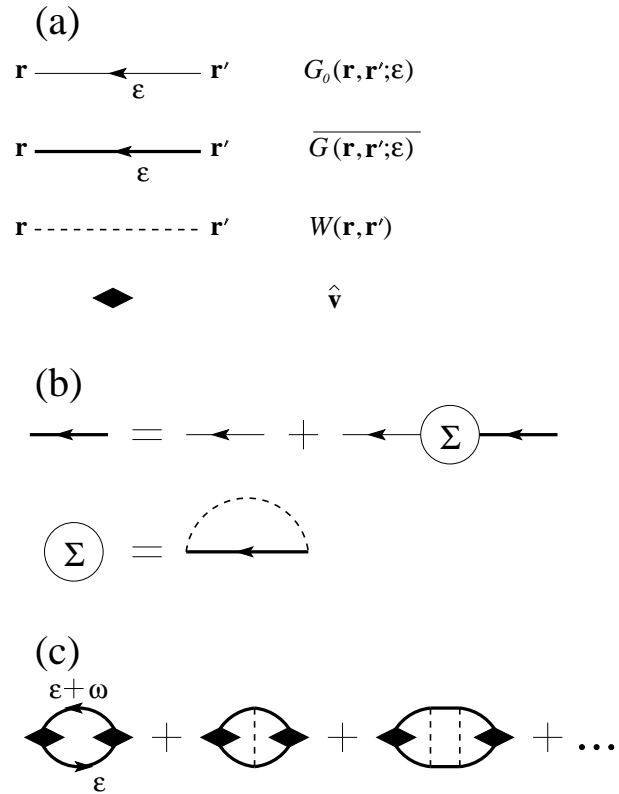}
\caption{\label{fig:SCBA}
(a)~Notations for different elements, constituting the diagrams.
(b)~diagrammatic representation of the SCBA equations
(\ref{GSCBA=}), (\ref{SigmaSCBA=}).
(c)~SCBA diagrams for the disorder-averaged conductivity.
SCBA, sometimes also called the non-crossing approximation,
corresponds to neglecting all diagrams with crossing dashed lines.
}
\end{figure}

In the presence of disorder, one can define the full Green's
function for each disorder realization,
\begin{equation}\begin{split}
{G}(\vec{r},\vec{r}';\ep)&=
\langle\vec{r}|(\ep-\hat{{H}})^{-1}|\vec{r}'\rangle
=\sum_{s}
\frac{\phi_s(\vec{r})\,\phi_s^*(\vec{r}')}{\ep-E_s},
\end{split}\end{equation}
($\phi_s$ and $E_s$ being the wave functions and energies of
the exact eigenstates of~$\hat{{H}}$),
as well as its average over the realization,
$\overline{{G}(\vec{r},\vec{r}';\ep)}$,
from which several physical quantities may be obtained, as will be
seen below. In the so-called self-consistent Born approximation
(SCBA), which can be represented diagrammatically using the standard
rules\cite{AGD} as shown in Fig.~\ref{fig:SCBA}(b),
one obtains the following system of closed equations for the average
Green's function~$\overline{{G}}$ and the
self-energy~$\Sigma$~\cite{AGD}:
\begin{subequations}\begin{eqnarray}
&&\overline{{G}(\vec{r},\vec{r}')}={G}_0(\vec{r},\vec{r}')
+\nonumber\\
&&\qquad{}+
\int{G}_0(\vec{r},\vec{r}_1)\Sigma(\vec{r}_1,\vec{r}_2)
\overline{{G}(\vec{r}_2,\vec{r}')}\,d\vec{r}_1d\vec{r}_2,
\label{GSCBA=}\\
&&\Sigma(\vec{r},\vec{r}')=W(\vec{r}-\vec{r}')
\overline{{G}(\vec{r},\vec{r}')}.
\label{SigmaSCBA=}
\end{eqnarray}\end{subequations}
The $\ep$ argument, common for all functions, has been omitted
for the sake of compactness.
SCBA in strong magnetic fields has been analyzed in
Ref.~\onlinecite{Ando1} for the parabolic dispersion,
$\ep(\vec{p})=p^2/(2m)$, and was shown to be valid for
high Landau levels\cite{Ando3}, in agreement with the general
rule\cite{AGD}: SCBA is valid when the electron motion between
successive scattering events is quasiclassical (i.~e., contains
many de Broglie wavelengths). Even when this condition
is fulfilled, the SCBA misses effects related to coherent
multiple-impurity scattering, but such effects are beyond the
scope of the present work.

Below we show that for the $\delta$-correlated disorder,
Eq.~(\ref{VV=}), and for strong magnetic fields, the SCBA
equations (\ref{GSCBA=}), (\ref{SigmaSCBA=})
are straightforwardly solved for any dispersion
law~$\ep(\vec{p})$. Indeed, since
$W(\vec{r}-\vec{r}')=W_0\delta(\vec{r}-\vec{r}')$, and
since $\overline{{G}(\vec{r},\vec{r};\ep)}$ does not depend
on~$\vec{r}$, Eq.~(\ref{SigmaSCBA=}) gives
$\Sigma(\vec{r},\vec{r}';\ep)=\Sigma(\ep)\,\delta(\vec{r}-\vec{r}')$.
Thus, $\Sigma(\vec{r},\vec{r}';\ep)$ has no matrix elements
between different Landau levels, so
\begin{equation}\label{barG=}
\overline{{G}(\vec{r},\vec{r}';\ep)}=
\sum_l\frac{\mathcal{P}_l(\vec{r},\vec{r}')}{\ep-\ep_l-\Sigma(\ep)}.
\end{equation}
For $\Sigma(\ep)$ we have the self-consistency equation,
which can be written as
\begin{equation}\label{SigmaE=sumn}
\Sigma(\ep)=\sum_l\frac{\gamma_B^2}{\ep-\ep_l-\Sigma(\ep)},
\quad\gamma_B^2=\frac{W_0}{2\pi\ell_B^2}
\end{equation}
where we used Eq.~(\ref{Pnrr=}).
It is convenient to express the factor $\gamma_B^2$
in terms of the cyclotron frequency
$\omega_c=(\ep_{l+1}-\ep_l)/\hbar$
and the zero-field scattering rate $1/\tau_{B=0}$, assuming
that both vary weakly with energy on the scale of $\hbar\omega_c$.
The self-consistent Born approximation at zero field
gives~\cite{AGD}
\begin{equation}
\frac\hbar{\tau_{B=0}}=2\pi\mathcal{N}{W}_0\nu,
\end{equation}
where $\mathcal{N}$ is the number of valleys (provided that
the disorder is sufficiently short-range to induce efficient
intervalley scattering), and $\nu$ is the density of states
per unit area per spin projection and per valley at zero field.
When the energy dependence of $\nu$ is weak, one can write
\begin{equation}
\nu\approx\frac{1}{\hbar\omega_c}\frac{1}{2\pi\ell_B^2},
\end{equation}
since each Landau level has $1/(2\pi\ell_B^2)$ states per unit
area, and $\hbar\omega_c$ is the separation between the Landau
levels. This gives
\begin{equation}
\gamma_B^2=\frac\hbar{\tau_{B=0}}\,
\frac{\hbar\omega_c}{2\pi\mathcal{N}}.
\end{equation}

In the limit of strong fields, $\omega_c\tau_{B=0}\gg{1}$,
we have $\gamma_B\ll\ep_l-\ep_{l'}$. Let us focus on some
Landau level~$l$ and on energies $E$ close to $\ep_l$.
First, let us consider terms in the sum in
Eq.~(\ref{SigmaE=sumn}) corresponding to levels different
from the chosen level~$l$. They result in a small overall
shift of the level~$l$:
\begin{equation}
\tilde\ep_l\approx\ep_l+\Sigma(\ep_l)\approx
\ep_l+\sum_{l'\neq{l}}\frac{\gamma_B^2}{\ep_l-\ep_{l'}}.
\end{equation}
In the following we will neglect the difference
$\tilde\ep_l-\ep_l\sim\gamma_B^2/(\hbar\omega_c)$, which is
much smaller than the Landau level broadening, as we will
see shortly. The term in the sum in Eq.~(\ref{SigmaE=sumn}),
corresponding to the same level~$l$, should be treated
exactly; it results in a quadratic equation for~$\Sigma(\ep)$.
As a result, we can write
\begin{equation}\label{Sigma=}
\Sigma(\ep)=\frac{\ep-\ep_l}2
-\sqrt{\frac{(\ep-\ep_l)^2}4-\gamma_B}
+\sum_{l'\neq{l}}\frac{\gamma_B^2}{\ep-\ep_{l'}},
\end{equation}
for $|\ep-\ep_l|\ll\hbar\omega_c$.

The knowledge of $\Sigma(\ep)$, Eq.~(\ref{Sigma=}),
and of $\overline{{G}(\vec{r},\vec{r}';\ep)}$,
Eq.~(\ref{barG=}), enables us to find the disorder-averaged
density of states per unit area per spin projection and per
valley:
\begin{subequations}\begin{eqnarray}
&&\overline{\nu(\ep)}=\frac{1}{\pi}\Im
\overline{{G}(\vec{r},\vec{r};\ep-i0^+)}
=\sum_l\frac{\rho(\ep-\ep_l)}{2\pi\ell_B^2},\qquad\\
&&\rho(\ep)=\frac{1}{\pi\gamma_B}
\sqrt{1-\left(\frac{\ep}{2\gamma_B}\right)^2}.
\end{eqnarray}\end{subequations}
Thus, each Landau level, which in the absence of disorder
is discrete and inifinitely degenerate, in the presence of
disorder is broadened into a semicircle.

To describe the cyclotron resonance (that is, inter-Landau-level
absorption), we start from the expression for two-dimensional
conductivity in terms of the exact wave functions
$\phi_s(\vec{r})$ and energies $E_s$, following directly from
the Kubo formula:
%\begin{equation}\begin{split}
%\sigma_{ij}(\omega)={}&\frac{2\mathcal{N}ie^2}{\omega}
%\int\frac{d^2\vec{r}}{L_xL_y}
%\sum_sf_s\,\phi_s^*(\vec{r})\,\hat{m}^{-1}_{ij}\phi_s(\vec{r})
%+{}\\
%&{}+\frac{2\mathcal{N}ie^2}\omega\,\frac{1}{L_xL_y}\sum_{ss'}
%\frac{f_{s}-f_{s'}}{\hbar\omega-(E_{s'}-E_s)+i0^+}\\
%&{}\times
%\int{d}^2\vec{r}\,\phi_{s}^*(\vec{r})\,\hat{v}_i\,\phi_{s'}(\vec{r})
%\int{d}^2\vec{r}'\,\phi_{s'}^*(\vec{r}')\,\hat{v}_j'\,\phi_{s}(\vec{r}')\\
%={}&\frac{2\mathcal{N}\hbar{e}^2}{iL_xL_y}\sum_{ss'}
%\frac{f_{s'}-f_{s}}{(E_{s'}-E_s)[\hbar\omega-(E_{s'}-E_s)+i0^+]}\\
%&{}\times
%\int{d}^2\vec{r}\,\phi_{s}^*(\vec{r})\,\hat{v}_i\,\phi_{s'}(\vec{r})
%\int{d}^2\vec{r}'\,\phi_{s'}^*(\vec{r}')\,\hat{v}_j'\,\phi_{s}(\vec{r}').
%\label{sigmaphis=}
%\end{split}\end{equation}
\begin{equation}\begin{split}
\sigma_{ij}^{(2)}(\omega)={}&\frac{2\mathcal{N}ie^2}{\omega{L}_xL_y}
\sum_sf_s\langle{s}|\hat{m}^{-1}_{ij}|s\rangle
+{}\\
&{}+\frac{2\mathcal{N}ie^2}{\omega{L}_xL_y}\sum_{ss'}
\frac{(f_{s}-f_{s'})\langle{s}|\hat{v}_i|s'\rangle
\langle{s'}|\hat{v}_j|s\rangle}{\hbar\omega-(E_{s'}-E_s)+i0^+}\\
={}&\frac{2\mathcal{N}i\hbar{e}^2}{L_xL_y}\sum_{ss'}
\frac{(f_{s}-f_{s'})\langle{s}|\hat{v}_i|s'\rangle
\langle{s'}|\hat{v}_j|s\rangle}{(E_{s'}-E_s)(\hbar\omega-E_{s'}+E_s+i0^+)}.
\label{sigmaphis=}
\end{split}\end{equation}
Here we used the notations
\begin{equation}
\hat{\vec{v}}=\left.\frac{\partial\ep(\vec{p})}{\partial\vec{p}}
\right|_{\vec{p}\to-i\hbar\vec\nabla},\quad
\hat{m}_{ij}^{-1}=\left.
\frac{\partial^2\ep(\vec{p})}{\partial{p}_i\partial{p}_j}
\right|_{\vec{p}\to-i\hbar\vec\nabla},
\end{equation}
%\begin{subequations}\begin{eqnarray}
%&&\hat{\vec{v}}=\left.\frac{\partial\ep(\vec{p})}{\partial\vec{p}}
%\right|_{\vec{p}\to-i\hbar\vec\nabla},\\
%&&\hat{m}_{ij}^{-1}=\left.
%\frac{\partial^2\ep(\vec{p})}{\partial{p}_i\partial{p}_j}
%\right|_{\vec{p}\to-i\hbar\vec\nabla},\\
%&&\langle{s'}|\hat{O}|s\rangle=
%\int{d}^2\vec{r}\,\phi_{s'}^*(\vec{r})\,\hat{O}\,\phi_{s}(\vec{r}),
%\end{eqnarray}\end{subequations}
and the last equality in Eq.~(\ref{sigmaphis=}) was obtained
using the commutation relations
\begin{equation}
\hat{m}_{ij}^{-1}=\frac{i}{\hbar}\left[\hat{v}_i,x_j\right],\quad
\hat{\vec{v}}=\frac{i}{\hbar}\left[\hat{H}_0,x_j\right].
\end{equation}
Introducing the notation
\begin{equation}
{G}^{R-A}(\vec{r},\vec{r}';\ep)=
G(\vec{r},\vec{r}';\ep+i0^+)-G(\vec{r},\vec{r}';\ep-i0^+),
\end{equation}
we can write the dissipative part of the conductivity,
averaged over the disorder, as
\begin{equation}\begin{split}
\overline{\Re\sigma_{xx}^{(2)}(\omega)}
={}&{}\frac{2\mathcal{N}e^2}{4\pi\omega}
\int{d}\ep\left[f(\ep+\hbar\omega)-f(\ep)\right]
\int\frac{{d}^2\vec{r}\,d^2\vec{r}'}{L_xL_y}
\\
&{}\times\overline{\hat{v}_x\,{G}^{R-A}(\vec{r},\vec{r}';\ep+\hbar\omega)\,
\hat{v}_x'\,{G}^{R-A}(\vec{r}',\vec{r};\ep)}.
\end{split}\end{equation}
This expression involves the average of products of two Green's
functions, which in SCBA reduces to summation of the ladder series,
shown in Fig.~\ref{fig:SCBA}(c). Consider the second term in this
series, containing one impurity line. It contains the spatial
integral
\[
\int{d}^2\vec{r}_1\,W_0\,J^{\pm,\pm}_{\ep,\ep+\omega}(\vec{r}_1)\,
J^{\pm,\pm}_{\ep+\omega,\ep}(\vec{r}_1),
\]
where $\vec{r}_1$ is the position of the impurity and
\begin{equation}
J^{\pm,\pm}_{\ep,\ep'}(\vec{r}_1)
=\int{d}^2\vec{r}\,\overline{G(\vec{r}_1,\vec{r};\ep\pm{i}0^+)}\,
\hat{v}_x\,\overline{G(\vec{r},\vec{r}_1;\ep\pm{i}0^+)}.
\end{equation}
Since the average Green's functions depend only on the relative
coordinate, $J^{\pm,\pm}_{\ep,\ep'}(\vec{r}_1)$ does not
depend on~$\vec{r}_1$, and thus it can be written as
\begin{equation}
J^{\pm,\pm}_{\ep,\ep'}
=\int\frac{{d}^2\vec{r}\,d^2\vec{r}_1}{L_xL_y}\,
\overline{G(\vec{r}_1,\vec{r};\ep\pm{i}0^+)}\,
\hat{v}_x\,\overline{G(\vec{r},\vec{r}_1;\ep\pm{i}0^+)}.
\end{equation}
Recalling Eq.~(\ref{barG=}) for the Green's function, we note that
because of the integration over $\vec{r}_1$, the contribution
to transitions between different Landau levels $l$ and $l'$
involves the product of the projectors
$\hat{\mathcal{P}}_l\hat{\mathcal{P}}_{l'}=0$. Thus, only the
first term of the ladder series in Fig.~\ref{fig:SCBA}(c)
contributes to the inter-Landau-level transition, in full
analogy with the case of the parabolic spectrum\cite{AndoJPSJ1975}.

\begin{figure}
\includegraphics[width=7cm]{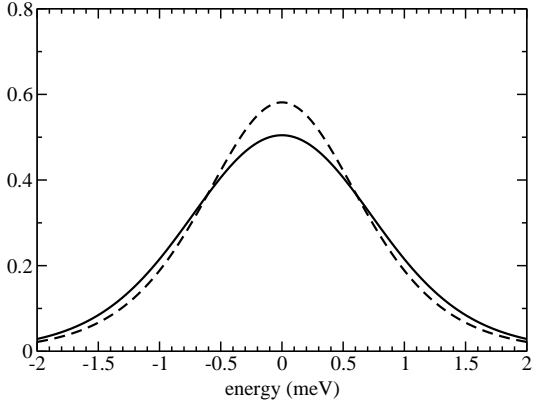}
\caption{\label{fig:Fermifunction}
Plot of $[f(\ep-\hbar\omega)-f(\ep+\hbar\omega)]/(\hbar\omega)$
(solid line)
and $-\partial{f}(\ep)/\partial\ep$ (dashed line)
as a function of $\ep-\ep_F$ for $T=0.43\:\mbox{meV}$,
$\hbar\omega=1.17\:\mbox{meV}$.}
\end{figure}

\begin{figure}
\vspace*{5mm}
\includegraphics[width=7cm]{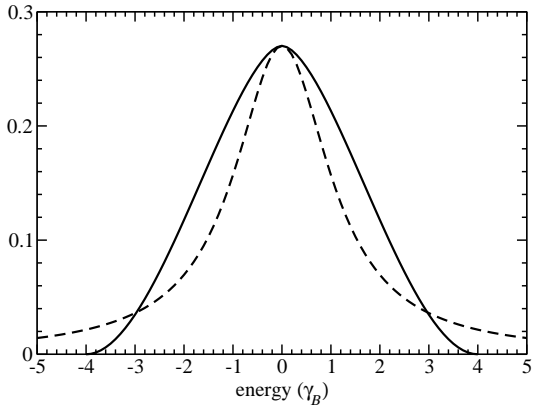}
\caption{\label{fig:semicircle}
Plot of the convolution
$\int\rho(\ep-\hbar\tilde\omega/2)\,
\rho(\ep+\hbar\tilde\omega/2)\,d\ep$
(solid line) and of the Lorentzian profile
$(\Gamma/\pi\hbar)/(\tilde\omega^2+\Gamma^2)$
(dashed line) as a function of $\hbar\tilde\omega/\gamma_B$
for $\Gamma=(3\pi/8)(\gamma_B/\hbar)$.
}
\end{figure}

Explicitly, from Eq.~(\ref{barG=}) we obtain
\begin{equation}\begin{split}
\overline{\Re\sigma_{xx}^{(2)}(\omega)}
={}&{}2\mathcal{N}\pi{e}^2
\int{d}\ep\,\frac{f(\ep-\hbar\omega/2)-f(\ep+\hbar\omega/2)}{\hbar\omega}\\
&{}\times
\sum_{l,l'}\hbar\,\rho(\ep-\hbar\omega/2-\ep_l)\,
\rho(\ep+\hbar\omega/2-\ep_{l'})
\\
&{}\times\sum_{\alpha,\alpha'}
\left|\langle{l,\alpha}|\hat{v}_x|l',\alpha'\rangle\right|^2.
\label{Resigmaxx=}
\end{split}\end{equation}
In this expression the disorder enters only in the densities
of states, $\rho(\ep)$, while the velosity matrix elements are
the same as in the absence of disorder. Moreover, when the
quasiclassical approximation is valid, ($\ep_l$~depends smoothly
on~$l$) they can be calculated classically. Indeed, comparing
Eqs. (\ref{sigmacl=}) and (\ref{Resigmaxx=}), one can establish
the correspondence
\begin{subequations}\begin{eqnarray}
-\frac{\partial{f}}{\partial\ep}&\leftrightarrow&
\frac{f(\ep-\hbar\omega/2)-f(\ep+\hbar\omega/2)}{\hbar\omega},\\
|v_{x,n}|^2&\leftrightarrow&\frac{2\pi\ell_B^2}{L_xL_y}
\sum_{\alpha,\alpha'}
\left|\langle{l,\alpha}|\hat{v}_x|l+n,\alpha'\rangle\right|^2,
\qquad\\
\frac{\Gamma/\pi}{\tilde\omega^2+\Gamma^2}
&\leftrightarrow&
\hbar\int\rho(\ep)\,\rho(\ep+\hbar\tilde\omega)\,d\ep,
\end{eqnarray}\end{subequations}
where we denoted $\tilde\omega=\omega-n\omega_c$ or
$\omega-(\ep_{l+n}-\ep_l)/\hbar$.
The first two lines hold reasonably well in our case
(even though
$T=5~\mbox{K}=0.43\:\mbox{meV}<\hbar\omega=1.17\:\mbox{meV}$,
see Fig.~\ref{fig:Fermifunction}),
while in the third one the convolution of two semicircles
is somewhat different from a Lorentzian, so there is no
unique relation between $\Gamma$ and $\gamma_B$.
We choose to match the peak heights (that is, the value
at $\tilde\omega=0$), which fixes
\begin{equation}
\Gamma=\frac{3\pi}{8}\frac{\gamma_B}{\hbar},
\end{equation}
as illustrated by Fig.~\ref{fig:semicircle}.

\end{document}